\documentclass[amsmath,amssymb,aps,prl,showpacs,floatfix,superscriptaddress,reprint,notitlepage]{revtex4-1}

\usepackage{amsmath,amssymb,textcomp,graphicx,mathtools}
\usepackage[colorlinks=true,linkcolor=blue,urlcolor=blue,citecolor=blue]{hyperref}
\usepackage[svgnames]{xcolor}
\usepackage{epstopdf}

\newcommand{\rr}{\mathbf{r}}
\newcommand{\q}{\mathbf{q}}
\newcommand*\diff{\mathop{}\!\mathrm{d}}

\begin{document}

\title{Binding energies of trions and biexcitons in two-dimensional
  semiconductors from diffusion quantum Monte Carlo calculations}

\author{M.\ Szyniszewski}

\affiliation{Department of Physics, Lancaster University, Lancaster LA1 4YB,
  United Kingdom}

\affiliation{National Graphene Institute, University of Manchester, Booth
  St E, Manchester M13 9PL, United Kingdom}

\author{E.\ Mostaani}

\affiliation{Department of Physics, Lancaster University, Lancaster LA1 4YB,
  United Kingdom}

\affiliation{Cambridge Graphene Centre, Engineering Department, University of
  Cambridge, 9 J.\ J.\ Thomson Avenue, Cambridge CB3 0FA, United Kingdom}

\author{N.\ D.\ Drummond}

\affiliation{Department of Physics, Lancaster University, Lancaster LA1 4YB,
  United Kingdom}

\author{V.\ I.\ Fal'ko}

\affiliation{Department of Physics, Lancaster University, Lancaster LA1 4YB,
  United Kingdom}

\affiliation{National Graphene Institute, University of Manchester, Booth
  St E, Manchester M13 9PL, United Kingdom}

\begin{abstract}
Excitonic effects play a particularly important role in the optoelectronic
behavior of two-dimensional (2D) semiconductors.  To facilitate the
interpretation of experimental photoabsorption and photoluminescence spectra
we provide statistically exact diffusion quantum Monte Carlo binding-energy
data for Mott-Wannier models of excitons, trions, and biexcitons in 2D
semiconductors.  We also provide contact pair densities to allow a description
of contact (exchange) interactions between charge carriers using first-order
perturbation theory. Our data indicate that the binding energy of a trion is
generally larger than that of a biexciton in 2D semiconductors.  We provide
interpolation formulas giving the binding energy and contact density of 2D
semiconductors as functions of the electron and hole effective masses and
the in-plane polarizability.
\end{abstract}


\maketitle

The optical properties of two-dimensional (2D) semiconductors such as
monolayer MoS$_2$, MoSe$_2$, WS$_2$, WSe$_2$, InSe, and phosphorene have
recently attracted a great deal of interest
\cite{2Doptical1,2Doptical2,2Doptical3,2Doptical4,2Doptical5,2Doptical6,2Doptical7,InSepaper}.
Numerous observations have been made of the rich structure of the luminescence
spectra of these 2D materials, in which the most pronounced features have been
interpreted in terms of neutral excitons
\cite{Excitons1,Excitons2,Excitons3,Excitons4,Excitons5,Excitons6}, charged
excitons (trions) \cite{Trions1,Trions2,Trions3,Trions4,Trions5,Trions6}, and
biexcitons \cite{Biexcitons1,Biexcitons2,Biexcitons3}, while recent
experiments on higher-quality monolayer transition-metal dichalcogenide (TMDC)
samples have revealed additional structure in their spectra
\cite{New1,New2,New3,New4}.

In this work we study a Mott-Wannier model of excitons and excitonic complexes
in monolayer 2D semiconductors, taking into account the polarizability of the
2D crystal \cite{Heinz1,Heinz2,Bogdan} and providing data to allow for a
perturbative treatment of contact interactions between carriers.  We use the
diffusion quantum Monte Carlo (DMC) approach
\cite{Ceperley1980,Foulkes2001,Needs2010} to find the energies
of trions and biexcitons, and we provide approximate formulas for
the exciton ($\cal U$), trion (${\cal E}_{\rm T}$), and biexciton (${\cal
  E}_{\rm XX}$) binding energies as functions of the in-plane polarizability
and the electron and hole effective masses, which fit the DMC data to within
5\%.  We calculate and report contact pair densities, enabling the evaluation
of perturbative corrections to the energies of charge-carrier complexes, as
well as intervalley scattering, due to contact (exchange) interactions between
charge carriers.  The strength of the contact interactions could in principle
be determined from first-principles calculations for different 2D
semiconductors; alternatively, the strengths of the contact interactions can
be regarded as parameters to be determined using experimental data in
conjunction with the contact pair densities reported here.

The energy $-{\cal U}-{\cal E}_{\rm T}$ of a trion with one hole (h) and two
electrons (e$_1$ and e$_2$) can be found by solving the Schr\"{o}dinger
equation (in Gaussian units)
\begin{eqnarray}
& & \left\{ \sum_{k={\rm e}_1,{\rm e}_2}
\left[-\frac{\hbar^2\nabla^2_{\rr_k}}{2m_{\rm e}} - U\left(\rr_{k{\rm
      h}}\right) \right]-\frac{\hbar^2\nabla_{\rr_{\rm h}}^2}{2m_{\rm h}}
+U(\rr_{{\rm e}_1{\rm e}_2}) \right\} \Psi \nonumber \\ & & \hspace{14em} {} =
\left[- {\cal U} - {\cal E}_{\rm T}\right] \Psi,
\label{eq:Schroedinger-trion}
\end{eqnarray}
where $m_{\rm e}$ and $m_{\rm h}$ are the electron and hole effective masses
and ${\bf r}_{ij}={\bf r}_i-{\bf r}_j$ is the position of particle $i$
relative to particle $j$.  The Keldysh potential $U$ describes the Coulomb
interaction screened by the polarization of the electron orbitals in the 2D
lattice \cite{Heinz1,Heinz2,Bogdan,Keldysh},
\begin{eqnarray}
U({\bf r}) & = & \frac{e^2}{\epsilon}\int \frac{\diff^2\q}{(2\pi)^2}\frac{2\pi
  e^{i\q\cdot\rr}}{q\left(1+qr_*\right)} \nonumber \\ & = & \frac{\pi
  e^2}{2\epsilon r_{\ast}} \left[ H_0 (r/r_\ast) - Y_0 (r/r_\ast) \right],
\label{eq:U}
\end{eqnarray}
where $r_*=2\pi\kappa_\bot$ is a parameter directly related to the in-plane
susceptibility $\kappa_\bot$ of the material, which has dimensions of length,
and $\epsilon$ is the average permittivity of the media on either side of the
2D semiconductor.  The potential is expressed in terms of a Struve function
$H_0$ and a Bessel function of the second kind $Y_0$.  Equation
(\ref{eq:Schroedinger-trion}) determines the main contribution ${\cal E}_{\rm
  T}$ towards the trion binding energy, which is counted from the exciton
binding energy $\cal U$.  Similar Schr\"{o}dinger equations can be written for
an exciton and a biexciton.

\begin{figure}
\begin{center}
\includegraphics[width=\columnwidth]{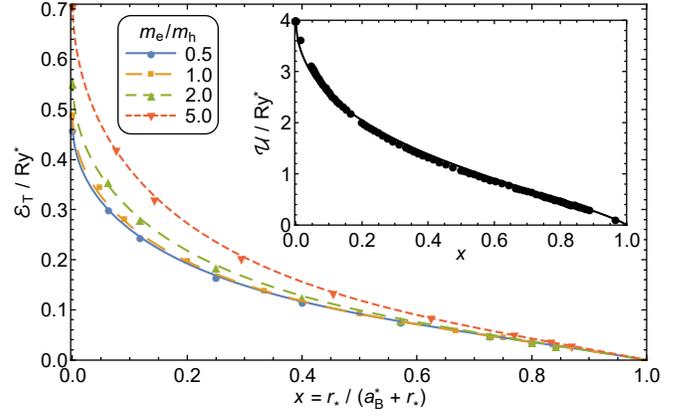}
\caption{(color online). Binding energies of trions at different mass ratios
  against rescaled in-plane polarizability $r_*$. The inset shows the binding
  energies of excitons against rescaled $r_*$. The lines show the fitting
  formulas of Eqs.\ (\ref{eq:Trion_binding_fit}) and
  (\ref{eq:Exciton_binding_fit}). \label{fig:trion_binding}}
\end{center}
\end{figure}

\begin{figure}
\begin{center}
\includegraphics[width=\columnwidth]{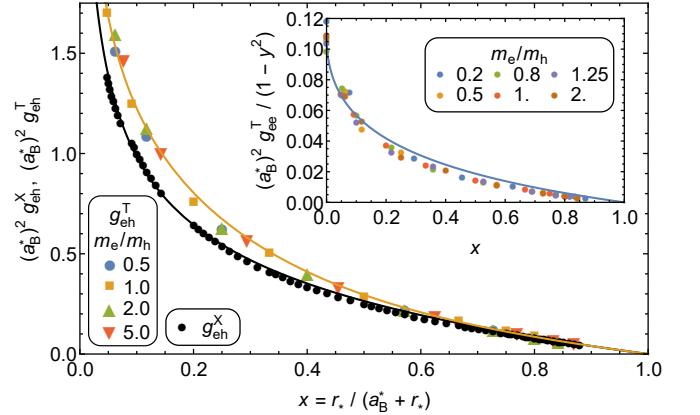}
\caption{(color online). Contact electron-hole pair densities of trions and
  excitons against rescaled in-plane polarizability $r_*$ at different mass
  ratios. The black curve is the fitting formula for an exciton,
  Eq.\ (\ref{eq:Exciton_g_fit}), while the yellow curve is the fitting formula
	for a trion, Eq.\ (\ref{eq:Trion_geh}).  The inset shows the (much smaller) contact
  pair density between the electrons in a negative trion and the fitting curve
  (in blue) of Eq.\ (\ref{eq:Trion_gee}).  \label{fig:trion_pair_density}}
\end{center}
\end{figure}

Numerical solution of the Mott-Wannier Schr\"{o}dinger equation for an exciton
yields the $r_*$-dependent binding energy ${\cal U}$ (see the inset in
Fig.\ \ref{fig:trion_binding}), which agrees with the asymptotic limits
\cite{Excitonlog1,Excitonlog2,Heinz1,Heinz2,Bogdan} ${\cal U}(r_*\rightarrow
\infty) \sim \frac{e^2}{\epsilon r_*}\ln \frac{r_*}{a_{\rm B}^*}$ and ${\cal
  U}(r_*\rightarrow 0)=-4\ R_{\rm y}^*$, as well as the contact pair density
$g^{\rm X}_{\rm eh}=\left< \delta({\bf r}_{\rm eh}) \right>$.  Their $r_*$ dependence
was fitted by
\begin{eqnarray}
{\cal U} / R_{\rm y}^* & \approx & (1-x)\left[ 4 - 1.0 x \ln{(1-x)}
  \right]/\left[1+1.31\sqrt{x} \right] \label{eq:Exciton_binding_fit} \\
g^{\rm X}_{\rm eh} & \approx & \frac{8.0}{(a_{\rm B}^*)^2} \frac{1-x}{1+20.0\sqrt{x}},
\label{eq:Exciton_g_fit}
\end{eqnarray}
where $x=r_*/(a_{\rm B}^*+r_*)$, $a_{\rm B}^*= \epsilon \hbar^2/(\mu
e^2)$ is the excitonic Bohr radius, $\mu=(m_{\rm e} m_{\rm h})/(m_{\rm e} +
m_{\rm h})$ is the reduced mass, and $R_{\rm y}^*= \mu e^4/(2 \epsilon^2 \hbar^2)$ is the excitonic Rydberg.  Here, the $\sqrt{x}$ term
is not a physical singularity; rather, it reflects the enhanced contact
density of a 2D hydrogen atom.  $g^{\rm X}_{\rm eh}$ is plotted in
Fig.\ \ref{fig:trion_pair_density}.

The ground-state solution to Eq.\ (\ref{eq:Schroedinger-trion}) for a trion
was obtained using the DMC approach \cite{Ceperley1980,Foulkes2001}, with the
trial wave function being optimized using variational Monte Carlo (VMC)\@.
The trial wave function was of the Jastrow form $\Psi = \exp [J ({\bf R})]$,
where the Jastrow exponent $J ({\bf R})$ consisted of a pairwise sum of terms
of the form $u_0(r) = \left[c_1 r^2 \log (r) + c_2 r^2 + c_3
  r^3\right]/\left(1 + c_4 r^2 \right)$, where $c_1$, $c_2$, $c_3$, and $c_4$
are optimizable parameters (different for each particle-pair type), together
with two-body and three-body polynomial terms that are truncated at finite
range \cite{Neil2004,Lopez2012}. The short-range behavior of $u_0(r)$ is such
that the analogs of the Kato cusp conditions for the logarithmic interaction
are satisfied (see the Supplemental Material in Ref.\ \cite{Bogdan}). Trial
wave functions were optimized by unreweighted variance minimization
\cite{Umrigar_1988,Drummond_2005} and energy minimization
\cite{Umrigar_2007}. The ground-state wave functions for these systems are
nodeless; hence the fixed-node DMC algorithm is exact.  The DMC calculations
were performed using the \textsc{casino} code \cite{Needs2010} with time steps
in the ratio $1 : 4$ and the corresponding target configuration populations in
the ratio $4 : 1$.  Afterwards, the energies were extrapolated linearly to
zero time step and hence, simultaneously, to infinite population.  The
resulting DMC trion binding energies, shown in Fig.\ \ref{fig:trion_binding},
agree with the asymptotic binding energies found earlier in the limits of
$r_*\rightarrow \infty$ \cite{Bogdan} and $r_*\rightarrow 0$ \cite{Spink2015}.
The trion contact pair densities $g^{\rm T}_{\rm ee}=\left< \delta({\bf
  r}_{{\rm e}_1{\rm e}_2}) \right>$ and $g^{\rm T}_{\rm eh}=\left< \delta({\bf
  r}_{{\rm e}_1{\rm h}}) + \delta({\bf r}_{{\rm e}_2{\rm h}}) \right>$ were
obtained by binning the radial distances sampled in the VMC and DMC
calculations, evaluating the extrapolated estimate of the pair density
\cite{Foulkes2001}, and then extrapolating the pair density to zero radial
separation. The resulting contact pair densities are shown in
Fig.\ \ref{fig:trion_pair_density}.  The trion binding-energy data are fitted
(to an accuracy within 5\%: see Fig.\ \ref{fig:trion_binding}) by the formula
\begin{eqnarray}
\frac{{\cal E}_{\rm T}}{R_{\rm y}^*} & \approx &
\left(1-\sqrt{x}\right) \Big[\left(0.44 x^2-1.16 \sqrt{x}+1.46\right) (2-y)
  \nonumber \\ & & \hspace{5em} {} -\left(0.64 x^2-2.0 \sqrt{x}+2.4\right)
\sqrt{1-y}\Big], \nonumber \\
\label{eq:Trion_binding_fit}
\end{eqnarray}
while the contact pair densities are fitted by
\begin{eqnarray}
g^{\rm T}_{\rm eh} & \approx & g^{\rm X}_{\rm eh}+ \frac{0.35}{(a_{\rm B}^*)^2}
(1-x)^{3.5} \quad {\rm and} \label{eq:Trion_geh} \\
g^{\rm T}_{\rm ee} & \approx & \frac{0.11}{(a_{\rm B}^*)^2}
\frac{1-\sqrt{x}}{1+\sqrt{x}}\left[1-y^2 \right],
\label{eq:Trion_gee}
\end{eqnarray}
where $y = \mu/m_{\rm h}$.  The term proportional to $\sqrt{1-y}$ in
Eq.\ (\ref{eq:Trion_binding_fit}) describes the contribution to the
ground-state energy due to the harmonic zero-point vibration of two heavy
electrons treated using the Born-Oppenheimer approximation \cite{Spink2015}.

\begin{figure}
\begin{center}
\includegraphics[width=\columnwidth]{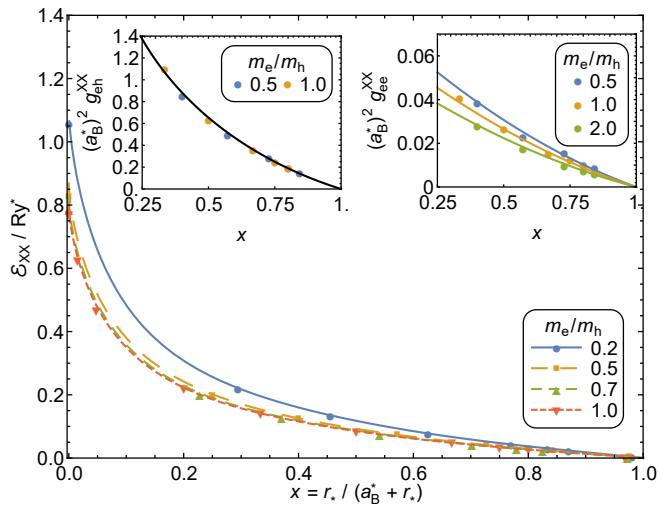}
\caption{(color online). Binding energies of biexcitons against rescaled
  in-plane polarizability $r_*$ at different mass ratios. The lines show the
  fitting formula of Eq.\ (\ref{eq:Biex_binding_fit}). The left inset shows
  the electron-hole contact pair densities for a biexciton and the approximation
	formula [black curve,
    Eq.\ (\ref{eq:Biexciton_geh})].  Electron-electron
  contact pair densities for a biexciton are shown in the right
  inset, together with the approximation formula of Eq.\ (\ref{eq:Biexciton_gee}).
	\label{fig:biex_binding}}
\end{center}
\end{figure}

Similarly, the binding energies ${\cal E}_{\rm XX}$ of biexcitons
were calculated using DMC and the results are presented in
Fig.\ \ref{fig:biex_binding}.  ${\cal E}_{\rm XX}$ is the energy required to
dissociate a biexciton into two separate excitons.  A fitting formula with up
to 5\% accuracy,
\begin{eqnarray}
\frac{{\cal E}_{\rm XX}}{R_{\rm y}^*} & \approx &
\left(1-\sqrt{x}\right)\left[1-1.2\sqrt{y(1-y)}\right] \nonumber \\ & & {}
\times\left[2.0-17.0x+43.0\left(x^{3/2}-x^2\right)+15.7x^{5/2}\right],
\nonumber \\
\label{eq:Biex_binding_fit}
\end{eqnarray}
incorporates the fact that the biexciton binding energy is symmetric under the
exchange of electrons and holes and includes the correct behavior in the
Born-Oppenheimer/harmonic-approximation limit of extreme mass ratio.
The biexciton
electron-hole and electron-electron contact pair densities can be approximated as
\begin{eqnarray}
g^{\rm XX}_{\rm eh} & \approx & 2 g^{\rm X}_{\rm eh}+ \frac{0.5}{(a_{\rm B}^*)^2}
(1-x)^2 \quad {\rm and} \label{eq:Biexciton_geh} \\
g^{\rm XX}_{\rm ee} & \approx & \frac{1-x}{(a_{\rm B}^*)^2}
(1 - 0.44 x ) (0.1 - 0.064 y).
\label{eq:Biexciton_gee}
\end{eqnarray}

\begin{figure}
\begin{center}
\includegraphics[width=\columnwidth]{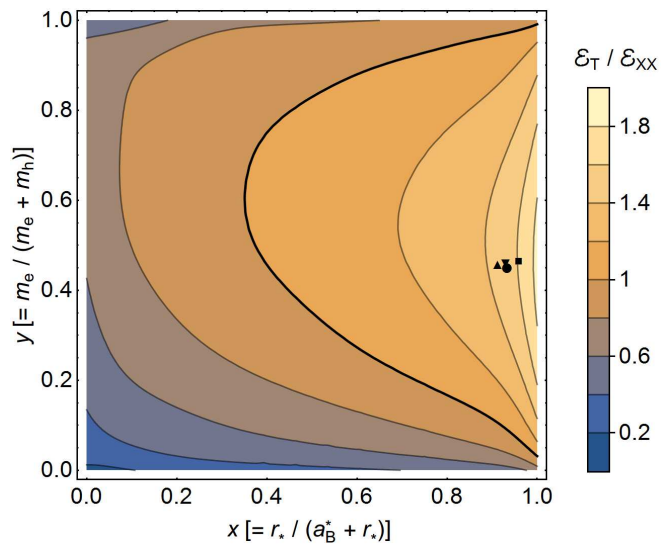}
\caption{(color online). Ratio of negative-trion binding energy to biexciton
  binding energy $({\cal E}_{\rm T}/{\cal E}_{\rm XX})$ as a function of
  rescaled in-plane polarizability $r_*$ and rescaled mass ratio. The thick
  black line shows the curve ${\cal E}_{\rm T}={\cal E}_{\rm
    XX}$. Experimentally relevant points for TMDCs are shown using symbols
  from Table\ \ref{table:example_binding_energies}.
	\label{fig:etrion-ebiex}}
\end{center}
\end{figure}

\begin{figure}
\begin{center}
\includegraphics[width=\columnwidth]{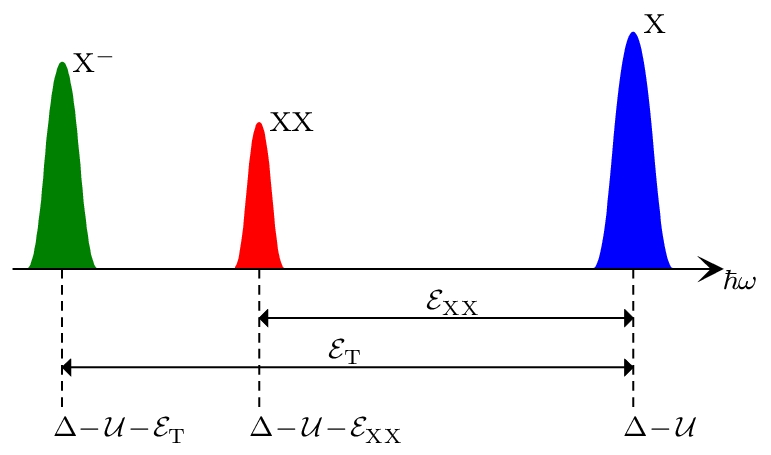}
\caption{(color online). Expected photoemission spectrum of a 2D semiconductor
  showing peaks for exciton (X), trion (X$^-$), and biexciton (XX)
  complexes. $\Delta$ is the quasiparticle band gap. \label{fig:spectrum}}
\end{center}
\end{figure}

The ratio of the negative-trion to the biexciton binding energy is
plotted against $x=r_*/(a_{\rm B}^*+r_*)$ and $y=\mu/m_{\rm h}=m_{\rm
  e}/(m_{\rm e}+m_{\rm h})$ in Fig.\ \ref{fig:etrion-ebiex}.  Although
the biexciton binding energy is larger than the trion binding energy
for the Coulomb interaction ($x=0$), the situation is generally
reversed when the interaction is of logarithmic form ($x=1$). However,
at extreme mass ratios, especially where the hole is heavy, the
biexciton is stabilized with respect to the negative trion.  In
practice 2D materials typically have $x>0.9$ and $y \approx 0.5$ (see
Table \ref{table:example_binding_energies}), and hence are strongly in
the regime in which the trion binding energy exceeds the biexciton
binding energy.  The qualitative form of our predicted trion spectrum
is shown in Fig.\ \ref{fig:spectrum}.  The trion peak occurs at lower
energy than the biexciton peak, in stark contradiction to the
classification of experimental peaks reported in
Ref.\ \cite{Plechinger_2015}. In fact several experimental works
\cite{Biexcitons1,Biexcitons2,Biexcitons3} have reported biexciton
binding energies of TMDCs that are about twice as large as the
reported trion binding energies
\cite{Trions1,Trions2,Trions3,Trions4,Trions5,Trions6}.
However the physical origins of experimentally
observed peaks in optical spectra are not always clear.  Our
conclusion that the trion binding energy is larger than the
biexciton binding energy is robust against large changes in the
values of the effective masses and the susceptibility and, taken at
face value, suggests that the experimental ``trion'' and
``biexciton'' peaks may be misclassified.

In Table \ref{table:example_binding_energies} we compare the trion and
biexciton binding energies obtained using Eqs.\ (\ref{eq:Trion_binding_fit})
and (\ref{eq:Biex_binding_fit}) with previous theoretical calculations and
experimental results in the literature for molybdenum and tungsten
dichalcogenide materials.  The theoretical results are in good agreement with
each other, and also with experimental results for the trion. However, for the
biexciton, there is a major disagreement between theory and experiment: the
experimental binding energies are around three times larger than the
theoretical biexciton binding energies.  Since our DMC solution of the
Mott-Wannier model is exact, the quantitative disagreement between the
positions of the theoretical and experimental trion and biexciton peaks must
indicate either a serious inaccuracy in the Keldysh interaction between charge
carriers or a misinterpretation of experimental spectra.  One possibility is
that contact (exchange) interactions between charge carriers as well as
intervalley scattering effects could play a significant role in charge-carrier
complexes.  Using our contact pair density data together with \textit{ab
  initio} calculations of contact interaction parameters could provide a
promising avenue for improving the quantitative description of the measured
photoemission spectra.

\begingroup
\begin{table*}
\caption{Trion and biexciton binding energies for an important group of TMDCs.
  We compare our fitting formulas with path integral Monte Carlo (PIMC)
  \cite{Kylanpaa_2015} and DMC data \cite{Mayers_2015} in the literature.
  Effective masses (in units of the bare electron mass) are taken from $GW$
  calculations in the literature
  \cite{Cheiwchanchamnangij_2012,Shi_2013}. \label{table:example_binding_energies}}
\begin{tabular}{clccccccccccc}
\hline \hline

& & & & & \multicolumn{4}{c}{Negative-trion binding energy ${\cal E}_{\rm T}$
  (meV)} & \multicolumn{4}{c}{Biexciton binding energy ${\cal E}_{\rm XX}$
  (meV)} \\

\multicolumn{2}{l}{\raisebox{1.5ex}[0pt]{Material}} &
\raisebox{1.5ex}[0pt]{$r_\ast$ ({\AA})} & \raisebox{1.5ex}[0pt]{$m_{\rm e}$} &
\raisebox{1.5ex}[0pt]{$m_{\rm h}$} & PIMC & DMC &
Eq.\ (\ref{eq:Trion_binding_fit}) & Exp.\ & PIMC & DMC &
Eq.\ (\ref{eq:Biex_binding_fit}) & Exp.\ \\

\hline

$\bullet$ & MoS$_2$ & $39$ \cite{Cheiwchanchamnangij_2012} & $0.35$ & $0.43$ &
$32.0$ & $33.8$ & $32$ & $34$ \cite{Trions2}, $35$ \cite{ZhangCleo} & $22.7(3)$
& $22.7(5)$ & $24$ & $60$ \cite{Sie_2015}, $70$ \cite{Biexcitons1} \\

${\scriptstyle \blacksquare}$ & MoSe$_2$ & $40$ \cite{Kumar_2012} & $0.38$ & $0.44$ & $27.7$ & $28.4$ & $31$ & $30$ \cite{Ross2013, Singh2014} & $19.3(5)$ & $17.7(3)$ & $23$ & N/A \\

$\blacktriangle$ & WS$_2$   & $38$ \cite{Berkelbach_2013} & $0.27$ & $0.32$ & $33.1$ & $34.0$ & $31$ & $34$ \cite{Zhu2015}, $36$ \cite{Heinz1} & $23.9(5)$ &
$23.3(3)$ & $23$ & $65$ \cite{Plechinger_2015} \\

$\blacktriangledown$ & WSe$_2$  & $45$ \cite{Berkelbach_2013} & $0.29$ & $0.34$ & $28.5$ & $29.5$ & $27$ & $30$ \cite{Trions3,Trions6} & $23.9(5)$ &
$23.3(3)$ & $20$ & $52$ \cite{Biexcitons3} \\

\hline \hline
\end{tabular}
\end{table*}
\endgroup

In summary we present exact numerical data for the ground-state solutions of
Mott-Wannier models of trions and biexcitons in 2D semiconductors in which the
charge carriers interact via the Keldysh interaction.  We have evaluated the
contact pair density between charge carriers, to permit subsequent
perturbative evaluations of the energy contribution due to contact exchange
interactions. Our results suggest that experimental spectra have been
misclassified, because the trion binding energy should exceed the biexciton
binding energy, but they also indicate that the Keldysh interaction fails to
give a quantitative description of the observed excitonic properties of 2D
TMDCs.  The contact pair density data that we provide will enable the
theoretical and experimental exploration of the role played by contact
interactions between charge carriers and intervalley scattering in 2D
semiconductors.

\begin{acknowledgments}
The authors thank T.\ Heinz, K.\ Novoselov, M.\ Potemski, A.\ Tartakovski, and
W.\ Yao for useful discussions.  This work was supported by EC FP7 Graphene
Flagship Project No.\ CNECT-ICT-604391, ERC Synergy Grant Hetero2D, EPSRC CDT
NOWNANO, and the Simons Foundation.  M.S.\ acknowledges financial support from
EPSRC, NOWNANO DTC grant number EP/G03737X/1. Computer resources were provided
by Lancaster University's High-End Computing cluster.
\end{acknowledgments}

\end{document}